\documentstyle[12pt,epsf]{ioplppt}
\begin{document}
\title{The effect of spin oscillation of relativistic particles
passing through substance and the possibility of the
constituent quark rescattering observation at
$ \Omega ^ {\hbox {-}} $-hyperon
 - proton collision
}[$ \Omega ^ {\hbox {-}} $-hyperon spin oscillation and quark rescattering]
\author{
 V G Baryshevsky, K G Batrakov and S Cherkas }
\address{
Nuclear Problems Institute,
Minsk, Belarus.}
\begin{abstract}

For the $ \Omega ^ {\hbox {-}} $-hyperon
passing through matter the phenomena
of spin rotation and oscillation has investigated
quantitatively.
Connection of these phenomena  with constituent quark rescattering
has been determined. It allows one to investigate quark rescattering
directly without background of a single scattering.
\end{abstract}
\section{ Introduction }

The research of spin effects in collisions of high energy particles
gives important information about their fundamental properties :
the quark-gluon interaction,  particle wave function, the
chiral symmetry violation mechanism, etc (Yndurain 1993).
In the local quantum field theory there are two fundamental principles:
unitarity and analyticity. It is known that unitarity links
the imaginary part of the forward scattering amplitude  with the total
cross section for colliding particles. In general case the
total cross section depends
on a spin orientation and  can be measured. The dispersion relations
between real and imaginary parts of the forward scattering amplitude
can be obtained from analyticity. By measuring imaginary and
real parts of a zero angle scattering amplitude  in a wide energy range
we get a possibility to check up unitarity and analicity principles.
As the direct measurements of the real part of the forward scattering
amplitude  in ordinary scattering experiments are practically impossible,
it is necessary to measure the differential cross section
and other observables for small-angle range of the coulomb-nuclear
interference with subsequent extrapolation of data to the zero angle
(Akchurin \etal 1993). There is, however, a possibility of the
direct measurement of the real part of the forward scattering amplitude
(Baryshevsky 1992, 1993). It was shown
that under penetration of a particle with
$ S \geq 1 $ spin through the medium two effects:
spin rotation and oscillation appear.
The magnitude of these effects is
determined by the real part of
the spin-dependent forward scattering amplitude. The unic
peculiarity of the spin oscillation phenomenon is that
  the  spin oscillation effect  does not decrease
with energy increasing, but even gains
(Baryshevsky 1993). This phenomenon was analyzed for a deuteron
(Baryshevsky 1993). It was shown, that the  spin
dependent part of the forward scattering amplitude measured with
the help of spin oscillation phenomenon is determined
only by effects of nucleon rescattering and gives the
information about the nucleon interaction at small distances.

In the present
work we consider the phenomena of oscillation
and rotation of $ \Omega ^ {\hbox {-}} $-hyperon spin.
For description of this effect we use the
Glauber eikonal approximation (Czyz and Maximon 1969).
It explore the idea that the matter densities
of the colliding particles determine the small angle scattering
cross section and follows from the geometry and probability
calculus. It turns out that  the effects of the spin rotation,
oscillation and dichroism is sensitive to the hadron inner structure
i.e. allow to check the hypotheses of the
constituent quarks. Picture of hadrons as being made of spatially
separated constituent quarks is not new. It  was considered on
phenomenological grounds many times
(Anisovich \etal 1985)
since the pioneer works of
Gell-Mann  1964. After some period of misunderstanding
it was clearly stated by Gell-Mann that "constituent" and "current"
quarks are completely different objects. The constituent quarks or
"valons" (Hwa 1980) appear as quasiparticles, that is as current quarks
and surrounding them clouds of gluons and quark-antiquark pairs.
Spin-dependent part of the zero angle elastic scattering
amplitude of
$ \Omega ^ {\hbox {-}} $-hyperon
at an unpolarized proton measured with the help of
spin oscillation and dichroism is determined by the
shadowing effects which strongly depends on the structure of
hadrons.

\section{ The rotation and oscillation phenomenon of
a $ \Omega ^{\hbox {-}} $-hyperon spin}
In accordance with Baryshevsky 1992,1993
the well-known formula (Lax 1951)
\begin{equation}
\label{ind_ref1}
\hat n = 1 + \frac {2\pi \rho} {k ^ 2} \hat F (0)~,
\end{equation}
for the refraction index of
a particle
inside the matter can be applied to the
non-zero spin particles.
Here $ \rho $ is the density of scatterers
in matter (the number of scatterers in $1~cm^3$ ),
$ k $ is the wave number of an  incident particle.
$ \hat F (0) $ is the zero-angle elastic scattering
amplitude, which is an operator in spin space of the
incident particle. The dependence of the amplitude
 $ \hat F (0) $ on the orientation of colliding particle
spins gives rise to quasioptic effects (spin rotation,
 spin oscillation and dichroism) under the passage of
a particle through the medium (Baryshevsky 1992,1993).

For particles with spins $ S \geq 1 $ (Baryshevsky 1992,1993)
the rotation of a spin arises even in passing inside
unpolarized targets.

Consider the propagation of
$ \Omega ^ {\hbox {-}} $-hyperon through the unpolarized
medium in detail. Let the spin state of a particle
being incident on a target
is characterized by an initial spin wave
function $ \psi _ 0 $. Then the spin wave function of the particle
in the target can be written as:
\begin{equation}
\label{psi}
\hat \psi (z) = \exp \{\i k\hat n z \} \hat \psi _ 0~.
\end{equation}
If the target is unpolarized, the scattering
amplitude  at the zero angle is determined  by hyperon spin
properties only:
\begin {equation}
\label{f0}
\hat F (0) = \frac{9 F_{1/2}(0)-F_{3/2}(0)}{8} +
\frac{F_{3/2}(0)-F_{1/2}(0)}{2}({\bf S n})^2~,
\end {equation}
where $ F_{3/2}(0) $ and
$ F_{1/2}(0) $ correspond to zero angle scattering amplitudes for the
$ \Omega^{\hbox {-}} $ hyperon with the spin projection $ S_z=3/2 $ and
$ S_z=1/2 $ to the
direction $\bf n$ of the incident particle momentum.

\Eref{f0} contains only term with square-law dependence
of spin
(we consider T-invariant interactions,
therefore the odd powers of spin in
the amplitude should be absent).
Denoting
 magnetic
quantum number through
$ m $  we obtain from equations (\ref{ind_ref1}),(\ref{f0}):
\begin {eqnarray}
\label{ref_in2}
n _ m = 1 + \frac {2\pi \rho} {k^2}F_m (0)~, \\
\bs
n _ m = n _ m ^ \prime + \i n _ m ^ {\prime\prime} \nonumber~,
\end {eqnarray}
where $ n _ m ^ \prime, n _ m ^ {\prime\prime} $ are
the real and imaginary parts of the refraction index
for the particle in an eigenstate with spin
operator projection $S_z=m$.

It follows from \eref{f0}, that the  states with quantum
numbers $ m $ and $ -m $ are described by the same refraction
indexes, however $ n _ {\pm 3/2} \neq n _ {\pm 1/2} $. Thus this
difference determines such effects as spin rotation and
oscillation.

In general case the hyperon spin wave function  at a medium entry
can be written as:
\begin{equation}
\label{psi0}
\hat\psi _ 0 = \{\mbox{a} \, \e ^ {-\i \,\delta_ {3/2}},
\mbox{b} \, \e ^ {-\i \,\delta _ {1/2}}, \mbox{c}
\, \e ^ {-\i \,\delta _{-1/2}},
\mbox{d} \, \e ^ {-\i \,\delta _ {-3/2}} \}.
\end{equation}
Using equations (\ref{psi}),(\ref{ref_in2}),(\ref{psi0}), it is
possible to find the expressions for  describing
polarization vector,  quadrupole (rank two) and octupole tensors
 of $\Omega ^{\hbox {-}} $-hyperon
as a function of a particle path length
inside the target (see appendix).
In particular, for the polarization vector we have the following
expressions:
\begin{equation}
\eqalign{
\fl < S_x>=\{2\mbox{b}\mbox{c} \cos ( \delta_{1/2} -  \delta_{-1/2})
  \e^{-2 n^{\prime\prime}_{1/2} kz} +
  \sqrt{3}\mbox{a}\mbox{b}
\e^{- (n^{\prime\prime}_{3/2} +  n^{\prime\prime}_{1/2}) kz}
\\ \times
   \cos ( \delta_{3/2} -  \delta_{1/2}
+(n^{\prime}_{3/2} - n^{\prime}_{1/2})kz) + {\sqrt{3}} \mbox{c}\mbox{d}
\\ \times
 \e^{-  (n^{\prime\prime}_{3/2}
 +  n^{\prime\prime}_{1/2}) kz}
   \cos ( \delta_{-1/2} -  \delta_{-3/2} -  (n^{\prime}_{3/2}-
      n^{\prime}_{1/2})kz)\}/|\psi|^2~,
\\    \bs
\fl <S_y>=\{-2\mbox{b}\mbox{c} \sin ( \delta_{1/2} -  \delta_{-1/2})
    \e^{-2 n^{\prime\prime}_{1/2} kz} -
  \sqrt{3} \mbox{a}\mbox{b} \e^
     {-  (n^{\prime\prime}_{3/2} +  n^{\prime\prime}_{1/2})kz}
\\ \times
   \sin ( \delta_{3/2} -  \delta_{1/2} +
(n^{\prime}_{3/2}-n^{\prime}_{1/2})kz)
- \sqrt{3}\mbox{c}\mbox{d}
\\  \times
   \e^{-  (n^{\prime\prime}_{3/2}  +  n^{\prime\prime}_{1/2})kz}
   \sin ( \delta_{-1/2} -  \delta_{-3/2} -  (n^{\prime}_{3/2}
- n^{\prime}_{1/2})kz)\}/|\psi|^2~,
\\   \bs
\fl <S_z>=\{\case32 (\,\mbox{a}^2 -
  \mbox{d}^2)\e^{-2\, n^{\prime\prime}_{3/2}\,kz} +
  \case12 (\,\mbox{b}^2 -
   \mbox{c}^2)\e^{-2\, n^{\prime\prime}_{1/2}\,kz}\}/|\psi|^2~,
}
\end{equation}
\[
|\psi|^2=(\mbox{a}^2 +
  \mbox{d}^2)\, \e^{-2\, n^{\prime\prime}_{3/2}\,kz} +
  (\mbox{b}^2 +
  \mbox{c}^2)\, \e^{-2\, n^{\prime\prime}_{1/2}\,kz},
\]
where $z$ is the particle way length inside the medium.
From these  expressions it follows, that in general
case the spin dynamics of $~ \Omega ^ {\hbox {-}}$ - hyperon is
characterized by a superposition of two rotations in clockwise and
counter-clockwise directions.  Let's consider some particular cases.
If the initial polarization vector is normal to the particle
momentum so, the initial populations and phases of
the states with the quantum numbers $ m $ are equal to the
ones for the states with quantum numbers $ -m $ then the
components $ < S_ y> $, $ < S_ z> $ will remain zero during
the whole time of particle
penetration through the
medium  and $ < S _ x > $ oscillates.
In the appendix it is shown, that the components of quadrupole and
octupole tensors of $ \Omega ^ {\hbox {-}} $-hyperon oscillate too.
If the initial polarization vector is directed at
the acute angle to the momentum direction  the
polarization vector motion look like
rotation. If the initial polarization vector
is directed at the obtuse angle to the momentum direction
the spin rotates in the opposite direction.

\section{Amplitude of elastic scattering}
Let the $ \Omega ^ {\hbox {-}} $-hyperon
passes through a hydrogen target. As it has been shown above the
phenomenon of spin oscillations is determined by the forward
scattering amplitude of $ \Omega ^ {\hbox {-}} $ on an
unpolarized particle. To evaluate hyperon-proton scattering
amplitude assume that the fundamental constituents making up
the colliding hadrons in an elastic  $\Omega ^{\hbox{-}}
p\rightarrow\Omega ^{\hbox{-}}p$
scattering
(presumably constituent quarks
or valons) can scatter with an elementary amplitude
$ f _ {ij} (s, t) $, where the subscripts refer to different
flavors ( Furget \etal 1990), $ s $ is a squared
energy of colliding quarks in their center-of-mass frame,
$t$ is the square of the 4-impulse transferred. At the elastic small
angle  scattering $t$ is expressed through the transverse
transferred
momentum
 $ {\bf q} $: $ t =-q ^ {2} $.In accordance with
the eikonal model the hyperon-proton scattering amplitude
is written down as (Czyz and Maximon 1969):
\begin {eqnarray}
\fl
F_{m}({\bf q})=\frac{\i}{2\pi}\int \d^{2} {\bf b} \e^{\i{\bf qb}}<m\mid 1-\prod_
{i,j=1}^{3}(1-\Gamma_{ij}
({\bf b} -\vec \eta_{i} +\vec \eta_{j}^{\prime}))\mid m>
\nonumber
\\ \lo=
\frac{\i}{2\pi}\int \d^{2} {\bf b} \e^{\i{\bf qb}}\int \d^{2}
\vec \eta_{1} \d^{2}\vec \eta_{2}
\d^{2}\vec \eta_{1}^{\prime} \d^{2}\vec \eta_{2}^{\prime}
\wp_{m}(\vec \eta_{1},\vec \eta_{2})
\wp^{\prime}(\vec \eta_{1}^{\prime},\vec \eta_{2}^{\prime})
\nonumber
\\  \times
(1-\prod_{i,j=1}^{3}(1-\Gamma_{ij}({\bf b} -\vec \eta_{i}
+\vec \eta_{j}^
{\prime})))~,
\label{Czyz}
\end{eqnarray}
where $ \vec \eta _ {i}~ (\vec \eta ^ {\prime} _ {i}) $ is quarks
transverse coordinates in a hyperon (proton) relative to the
hadron center-of-mass: $ \vec \eta _ 3 = -\vec \eta _ 2- \vec \eta
_ 1, ~ \vec \eta _ 3 ^ \prime = -\vec \eta _ 2 ^ \prime-\vec \eta _
1 ^ \prime $. $ \wp _ {m} (\vec \eta _ {1}, \vec \eta _ {2}) $,~
$ \wp ^{\prime} (\vec \eta^{\prime} _ {1},\vec \eta^{\prime} _ {2})$
are quark distribution functions over transverse
coordinates in a hyperon and a proton respectively, $ m $ is the
projection of a $ \Omega ^{\hbox {-}} $-hyperon spin,
$ \Gamma _ {ij} ({\bf b}) $ is the profile - function connected
with the quark scattering amplitude $ f _ {ij} ({\bf q}) $ by
means of the following relation:
\begin {equation}
\Gamma_{ij}({\bf b})= {1\over{2\pi\i}}\int f_{ij}({\bf q})
\e^{-\i{\bf qb}} \d^{2}{\bf q}.
\end{equation}
The amplitude $ F _ {m} ({\bf q}) $ is normalized by the condition
( Furget \etal 1990):
\begin {equation}
{d\sigma \over dt}^{m} =\pi \mid F_{m}(q)\mid ^{2} ,\qquad
\sigma _{tot}^m = 4\pi Im F_{m}(0)~.
\end{equation}
In such a normalization the amplitude of scattering is invariant
relative to the Lorentz
transform along the incident particle momentum direction.
It can be obtained from the usual amplitude used in the formula
(\ref{ind_ref1}) through division by a wave number $ k $ (small-angle
scattering is meant). Note that the quark distribution function over
transverse coordinates is also invariant under this transforms.
This allows us to calculate it in the particle rest frame, by integrating
the ordinary distribution function $ \Re _ {m} ({\bf r} _ {1},
{\bf r} _ {2}) $ over all $ z_i $ coordinates, where
$ {\bf r} _ {i} \equiv (\vec \eta _ {i}, z _ {i}) $ and
$z$-axes parallel to the incident particle momentum direction.

For simplicity we consider the amplitude quark
scattering as spinless, therefore the  hyperon spin dependence
is reduced to the spin dependence of quark
distribution function $ \wp _ {m} (\vec \eta _ {1}, \vec \eta _ {2}) $.
Let's show, however, that for an unpolarized proton
the contribution of single quark
scattering process to the $F_m(0)$ amplitude does not depend
on a spin state of a hyperon $ m $ even
in the case of spindependent quark scattering
amplitude. Let's rewrite  formula (7)
for single scattering but taking into account quark spins:
\begin{eqnarray}
\fl
F^{(1)}_{m}(0)=
\frac{\i}{2\pi}\int \d^{2} {\bf b} \int \d^{2}\vec \eta_{1}..
\d^{2}\vec \eta_{3}
\d^{2}\vec \eta_{1}^{\prime}..\d^{2}\vec \eta_{3}^{\prime}
\Tr_{_{\stackrel{1^\prime2^\prime3^\prime}{123}}}[
\wp^{\prime}(\vec \eta_{1}^{\prime},\vec \eta_{2}^{\prime}
,\vec \eta_{3}^{\prime})
\wp_{m}(\vec \eta_{1},\vec \eta_{2},\vec \eta_{3})
\nonumber
\\ \ms \times
\sum_{i,j=1}^{3}\Gamma_{ij}({\bf b} -\vec \eta_{i}
+\vec \eta_{j}^{\prime})]=
\frac{\i}{2\pi}\sum_{i,j=1}^{3}
\int \d^{2} {\bf b} \int \d^{2}\vec \eta_{i}
\d^{2}\vec \eta_{j}^{\prime}\Tr_{_{\stackrel{i}{j}}}[
\wp_{m}(i,\vec \eta_{i})
\nonumber
\\
\ms \times
\wp^{\prime}(j,\vec \eta_{j}^{\prime})
\Gamma_{ij}({\bf b} -\vec \eta_{i} +\vec \eta_{j}^{\prime})]=
\sum_{i,j=1}^{3}\Tr_{_{\stackrel{i}{j}}}[
\wp_{m}(i)\wp^{\prime}(j)f_{ij}(0)].
\end{eqnarray}
The profile-function $ \Gamma _ {ij} ({\bf b}) $ and,
consequently the
amplitude $ f _ {ij} ({\bf q}) $ are the operators in the quark spin
space. $ \wp ^ {\prime} (\vec \eta _ {1} ^ {\prime},
\vec \eta _ {2} ^ {\prime}, \vec \eta _ {3} ^ {\prime}) $
and $ \wp _ {m} (\vec \eta _ {1}, \vec \eta _ {2}, \vec \eta _ {3}) $
represent simultaneously distribution functions of quarks over transverse
coordinates and spin density matrixes of quarks in a unpolarized proton
and hyperon with a projection of a spin $ m $ respectively. To consider
all quarks in the same way we  have formally included the integration
over third quark coordinates by putting
$
\wp_{m}(\vec \eta_{1},\vec \eta_{2},\vec \eta_{3})=
\wp_{m}(\vec \eta_{1},\vec \eta_{2})\delta(\vec \eta_{1}+
\vec \eta_{2}+\vec \eta_{3}).
$
\noindent
The forward scattering amplitude in case of P,T-conservation
has the following form:
\begin {equation}
f_{ij}(0)=f_{ij}^\prime + f_{ij}^{\prime\prime}\vec\sigma_i\vec
\sigma^\prime _j +
f_{ij}^{\prime\prime\prime}(\vec\sigma_i{\bf n})(\vec\sigma^\prime_j
{\bf n}),
\end{equation}
where $ \vec\sigma _ i $ and $\vec \sigma^\prime _ j $ are the spin
matrixes of hyperon and proton quarks, respectively.
The vector $ {\bf n} $ determines the direction of a hyperon motion
and we assume it to coincide with the direction of hyperons quark motion.
The averaging  $ \vec\sigma^\prime _ j $ over a spin density matrix
of quark in an unpolarized proton gives zero. Taking into
account  that $ Sp _i[\wp _{m}(i))=1$ and $ Sp _ j [\wp _ {m} ^
\prime (j)] = 1 $ we obtain that
\[
 F ^ {(1)} _ {m} (0) = \sum _ {i, j =
1} ^ {3} f ^ {\prime} _ {ij}~,
\]
and consequently does not depend on the spin state of a hyperon.
Everywhere further we consider quarks as spinless objects.
We shall not distinguish $ u $ and $ d $ quarks, as a result
the single scattering amplitude $ f (q) $ of $s$-quark on $ u $ or $ d $
quarks should be considered.

\section{The quark transverse coordinates distribution function}
As it has been already mentioned the elastic forward scattering amplitude
of $ \Omega ^{\hbox {-}} $ on an unpolarized proton depends on a
spin projection of $ \Omega ^{\hbox {-}} $ . Below we shall find
a quark transverse coordinates distribution function
in a hyperon for the states with the $ 3/2 $ and $ 1/2 $
spin projections relative to the momentum direction.
Considering $ \Re ({\bf r} _ {1},
{\bf r} _ {2}) $ as an operator in hyperon spin space
it is possible to expand it in terms of Cartesian spin-tensors
of $ \Omega ^ {\hbox{-}} $ - hyperon:
\begin {eqnarray}
\Re({\bf r}_{1},{\bf r}_{2})=\Re^{0}({\bf r}_{1},{\bf r}_{2})
+{1\over 2}\sum_{i,j}(S_{i}S_{j}+S_{j}S_{i}-\frac{5}{2}\delta_{ij})
\nonumber
\\
\times
(A(r_{1i}r_{1j}+ r_{2j}r_{2i})+Br_{1i}r_{2j})\stackrel{\sim}
\Re({\bf r}_{1},{\bf r}_{2}).
\end{eqnarray}
$ {\bf S} $ is the hyperon spin operator. In (12) there are not terms
of the first and third order of spin due to $ T $-invariance,
terms with higher spin powers  are omitted
by virtue of commutating
relations. Considering the matrix element of $ \Re ({\bf r} _ {1},
{\bf r} _ {2}) $ between the states  with  the spin projections $m$ we get:
\begin {eqnarray}
\Re_{m}({\bf r}_{1},{\bf r}_{2})=\Re^{0}({\bf r}_{1},{\bf r}_{2})
+(-1)^{m+1/2}\biggl(A(z_{1}^2+z_2^2-
\frac{\eta_1 ^2+\eta_2^2}{2})
\nonumber
\\
+
B(z_1z_2-\frac{\vec \eta_1\vec \eta_2}{2})\biggr)\stackrel{\sim}
\Re({\bf r}_{1},{\bf r}_{2})).
\end{eqnarray}
To find the functions $ \Re ^ {0} ({\bf r} _ {1}, {\bf r} _ {2}) $,
$ \stackrel {\sim} \Re ({\bf r} _ {1}, {\bf r} _ {2}) $
and the factors $ A, B $ it is necessary to choose some model
of $ \Omega ^{\hbox {-}} $-hyperon,
for example the nonrelativistic quark model (Gershtein, Zinoviev 1981).
The wave function for the spin projection 3/2 has been
calculated by Gershtein, Zinoviev 1981
and has the form:
\begin {equation}
\mid 3/2 >= 0.997\mid ^{4}S> - 0.068\mid ^{4}D> - 0.045\mid ^{2}D>.
\label{f32}
\end{equation}
\noindent
The expressions for $ \mid ^ {4} S>,~ \mid ^ {4} D>,~ \mid ^ {2} D> $
can be found in (Gershtein, Zinoviev 1981).
To simplify our consideration we omit
a small impurity  of symmetrical radial excitation
$ \mid ^ {4} S' > $ in \eref{f32}.
Multiplying the wave function of the $ \mid 3/2> $ state by
the function conjugate to it
the quark coordinates distribution function may be obtained:
\begin {eqnarray}
\fl
\Re_{3/2}({\bf r}_{1},{\bf r}_{2})=\frac{3\sqrt{3}}{R^{6}\pi^{3}}
\exp{\left(2\frac{-r_1^2-r_2^2-{\bf r_1r_2}}{R^2}\right)}\biggl[1
\nonumber
\\
~~~~~~~~~~~~~~~~~~~~+\frac{4
\times 0.068}{\sqrt{30}R^2}
(\eta_1^2+\eta_2^2+\vec \eta_1\vec\eta_2
-2(z_1^2+z_2^2+z_1z_2)\biggr]~,
\end{eqnarray}
\noindent where ${\bf r} _ {i} =
(\vec \eta _ {i}, z _ {i}) $.
Comparison of (15) with
(13)
gives $ A = B = -8\times 0.068/(\sqrt{30}R^{2}) $. $R ^{2} $
is a mean squared radius
of $ \Omega ^ {\hbox{-}} $-hyperon. As a result of integrating
equation (15) over $ z $ - coordinates,
the quark distribution function over transverse
coordinates turns out to be:
\begin {eqnarray}
\fl
\wp_{_{\stackrel{3/2}{1/2}}}(\vec \eta_1,\vec \eta_2)=\frac{3}
{R^4 \pi^2}\exp{\left(2\frac{
-\eta_1^2-\eta_2^2-\vec \eta_1\vec\eta_2
}{R^2}\right)}
\biggl[1\mp \frac{4\times 0.068}{\sqrt{30}R^2}\biggl(R^2
\nonumber
\\
~~~~~~~~~~~~~~~~~~~-
\eta_1^2-\eta_2^2-\vec \eta_1\vec\eta_2\biggr)\biggr].
\end{eqnarray}
At last we find a hyperon form factor required hereinafter:
\begin{eqnarray}
G_{_{\stackrel{3/2}{1/2}}}({\bf q,q}^{\prime})=\int
\wp_{_{\stackrel{3/2}{1/2}}}(\vec \eta_1,\vec \eta_2)\exp{(\i{\bf q}
\vec \eta_1
+\i{{\bf q}^\prime \vec \eta}_2)}\d^2\vec \eta_1\d^2\vec \eta_2
\nonumber
\\ \bs
=\exp{\left(\frac{(-q^2-q^{\prime 2}+{\bf qq}^{\prime})R^2}{6}\right)}
\left[1\mp\frac{Q}{12}(q^2+q^{\prime 2}-{\bf qq}^{\prime})\right],
\label{form-fac}
\end{eqnarray}
where $ Q = 8\times 0.068 R ^ 2/\sqrt {30} $
is the quadrupole moment of $ \Omega ^ {\hbox{-}} $-hyperon.
In the considered model of $ \Omega ^ {\hbox{-}} $-hyperon the
difference between form factors for the states with
the  $ \pm 3/2 $ and $ \pm 1/2 $ spin projections
is mainly determined mainly by
the impurity of $ \mid ^ {4} D> $ state, as well as by the quadrupole
moment:
\[
Q=\frac{2}{3}\int(
\eta_1^2+\eta_2^2+\vec \eta_1\vec\eta_2
-2(z_1^2+z_2^2+z_1z_2)
) \Re_{3/2}({\bf r}_{1},{\bf r}_{2}) \d^{3}
{\bf r}_{1}\d^{3}{\bf r}_{2}.
\]

\section{Expansion of the eikonal function on
scattering multiplicities}
The expression (\ref{Czyz}) for the scattering amplitude
can be rewritten as:
\begin {equation}
F_m ({\bf q})=\frac{\i}{2\pi}\int \d^2{\bf b}
\e^{\i{\bf qb}}(1-\gamma_m({\bf b}));
\label{fmm}
\end{equation}
\begin{eqnarray}
\gamma_m({\bf b})=<m\mid\prod_{ij=1}^3(1-\Gamma ({\bf b}-\vec \eta_i
+\vec \eta
^\prime_j)\mid m>
\nonumber
\\
=<m\mid 1-\sum_{ij}\Gamma_{ij}+\frac{1}{2!}\sum_{\stackrel{ij}{kl}}
^\prime
\Gamma_{ij}\Gamma_{kl}-\frac{1}{3!}\sum_{\stackrel{ijk}{lpn}}
^\prime
\Gamma_{ij}\Gamma_{kl}\Gamma_{pn}+...\mid m>.
\label{gammb}
\end{eqnarray}
The prime in the above sum means a lack of terms with pairwise
equal indexes in it, for example $ i = k $ and $ j = l $.
We limit ourself by the exact
account of double collisions and carry factorization
of the equation (19):
\begin{eqnarray}
\fl
\gamma_m({\bf b})\approx 1-<m\mid \sum_{i,j}\Gamma_{ij}\mid m>+
\frac{1}{2!}<m\mid \sum_{\stackrel{ij}{kl}}'\Gamma_{ij}\Gamma_{kl}\mid
m>
\nonumber
\\
-\frac{1}{3!}<m\mid \sum_{\stackrel{ij}{kl}}'\Gamma_{ij}\Gamma_{kl}\mid m>
<m\mid \sum_{pn}\Gamma_{pn}\mid m>
\nonumber
\\
+\frac{1}{4!}
<m\mid \sum_{\stackrel{ij}{kl}}'\Gamma_{ij}\Gamma_{kl}\mid m>^2-...
\nonumber
\\
=
ch\{\Omega^\prime_m({\bf b})\}-\Omega_m({\bf b})
\frac{sh\{\Omega^\prime_m({\bf b})\}}
{\Omega^\prime_m({\bf b})},
\label{chsh}
\end{eqnarray}
where
\numparts
\begin{eqnarray}
\fl
\Omega_m({\bf b})\equiv  <m\mid \sum_{i,j}\Gamma_{ij}\mid m>
=\frac{9}{2\pi\i}\int f({\bf q})G_m({\bf q})G^\prime ({\bf q})
\e^{-\i{\bf qb}}\d^2{\bf q}~,
\label{om1}
\\
\fl
(\Omega^\prime_m({\bf b}))^2\equiv <m\mid
\sum_{\stackrel{ij}{kl}}'\Gamma_{ij}\Gamma_{kl}\mid m>=
-\frac{18}{4\pi^2}\int f({\bf q})
f({\bf q}^\prime)[G^\prime ({\bf q+q}^\prime)
G_m ({\bf q,q}^\prime)
\nonumber
\\ +
G^\prime ({\bf q,q}^\prime)G_m ({\bf q+q}
^\prime)+2G_m({\bf q,q}^\prime) G^\prime
({\bf q,q}^\prime)]
\e^{-i({\bf q+q}^\prime){\bf b}}\d^2{\bf q}\d^2{\bf q}^\prime.
\label{om2}
\end{eqnarray}
\endnumparts
The form factor $ G _ {m} ({\bf q, q}^\prime ) $ has been defined
earlier (see \eref{form-fac})). $ G ^\prime({\bf q, q}^\prime) $
is a same quantity for a proton, $ G^\prime ({\bf q}) =
G^\prime  ({\bf q}, 0),~
G_{m}({\bf q})=G_m({\bf q},0)$.
The factorization (\ref{chsh}) is minimal, in the sense that the
 each item in \eref{gammb} is divided in the smallest number
of multiplicands. Another factorization is offered
by Franco,Varma 1978:
\begin {equation}
\gamma_m ({\bf b})\approx \exp\left(-\Omega_m
({\bf b})+{1\over 2} \left(\Omega_m
^{\prime}({\bf b})\right)^2-{1\over 2}\left(\Omega_m
({\bf b})\right)^2\right).
\label{exp}
\end{equation}
The total factorization of a series (19) corresponds to
\begin {equation}
\gamma_m({\bf b})\approx \exp\left(-\Omega_m({\bf b})\right)
~
\label{opt}
\end{equation}
i.e. to the optical limit.
The account only single and double collisions gives:
\begin {equation}
\gamma_m({\bf b})\approx 1- \Omega_m
({\bf b}) + {1\over 2}\left(\Omega_m^\prime({\bf b})\right)^{2}~.
\label{double}
\end{equation}
Let us show once more that
the single collision contribution to the
scattering amplitude at the zero angle does not depend on $m$:
$ F _ m ^ {(1)} (0) =
\frac {\i} {2\pi} \int \d ^ 2{\bf b}\Omega _ m ({\bf b})
= 9f (0) G_m(0)G^\prime(0)=9f (0) $.
Therefore the single scattering falls out
from the difference of forward scattering amplitudes
for the states with
different $ m $.
For further calculations the quark
scattering amplitude can be taken in the form:
\begin {equation}
\eqalign{
f({\bf q})=f(0)\exp{(-aq^2/2)},
\\  \bs
f(0)=\frac{(\i+0.2)\sigma_{qs}}{4\pi}.
}
\label{ami}
\end{equation}
$ \sigma _ {qs} = 3.2~mb $ is
the  cross
section of the
$s$-quark - $u,d$-quark scattering, $ a = 0.8 ~ Gev ^ {-2} $
(Furget \etal 1990), for the $292.4~ Gev$
 hyperon energy in the proton rest frame  ($ \sqrt {s} = 23.5 ~ Gev$).
Substituting the form factors (\ref{form-fac}) and amplitude
(\ref{ami}) to
the equations (\ref{om1}),(\ref{om2}) we obtain:
\numparts
\begin{eqnarray}
\fl
\Omega_{_{\stackrel{3/2}{1/2}}}({\bf b})=-9\i f(0)
\exp{\left(\frac{-b^2}{2a+\frac{2}{3}R^2+\frac{2}{3}R^{\prime 2}}
\right)}\left(a+\frac{1}{3}R^2+\frac{1}{3}R^{\prime 2}\right)^{-1}
\nonumber
\\ \ms
\times
\left[1\pm \frac{Q}{12}\left(\frac{b^2}
{(a+\frac{1}{3}R^2+\frac{1}{3}R^{\prime 2})^2}-
\frac{2}{a+\frac{1}{3}R^2+\frac{1}{3}R^{\prime 2}}\right)\right],
\label{om11}
\\ \ms \bs
\fl
\Omega^{\prime 2}_{_{\stackrel{3/2}{1/2}}}({\bf b})=
-18f^2(0)\Biggl\{
\exp{\left(\frac{-b^2}{a+\frac{2}{3}R^{\prime 2}+\frac{1}{6}R^2}
\right)}(a+\frac{1}{2}R^2)^{-1}
(a+\frac{2}{3}R^{\prime 2}+\frac{1}{6}R^2)^{-1}
\nonumber
\\  \ms  \bs
\fl ~~~~~~
\times
\Biggl[1\pm \frac{Q}{12}\left(\frac{b^2}
{(a+\frac{2}{3}R^{\prime 2}+\frac{1}{6}R^2)^2}
-
\frac{4a+2R^{\prime 2}+R^2}{(a+\frac{1}{2}R^2)
(a+\frac{2}{3}R^{\prime 2}+\frac{1}{6}R^2)}\right)\Biggr]
\nonumber
\\  \ms  \bs
+
\exp{\left(\frac{-b^2}{a+\frac{2}{3}R^2+\frac{1}{6}R^{\prime 2}}
\right)}
(a+\frac{1}{2}R^{\prime 2})^{-1}
(a+\frac{2}{3}R^2+\frac{1}{6}R^{\prime 2})^{-1}
\nonumber
\\  \ms \bs
\times
\left[1\pm \frac{Q}{3}\left(\frac{b^2}
{(a+\frac{2}{3}R^2+\frac{1}{6}R^{\prime 2})^2}-
\frac{1}{a+\frac{2}{3}R^2+\frac{1}{6}R^{\prime 2}}\right)
\right]
\nonumber
\\  \ms  \bs
+2
\exp{\left(\frac{-b^2}{a+\frac{R^2}{6}+\frac{R^{\prime 2}}{6}}
\right)}(a+\frac{R^2}{2}+\frac{R^{\prime 2}}{2})^{-1}
(a+\frac{R^2}{6}+\frac{R^{\prime 2}}{6})^{-1}
\nonumber
\\   \ms  \bs
\times
\left[1\pm \frac{Q}{12}\left(\frac{b^2}
{(a+\frac{R^2}{6}+\frac{R^{\prime 2}}{6})^2}-
\frac{4a+R^2+R^{\prime 2}}{(a+\frac{R^2}{6}+\frac{R^{\prime 2}}{6})
(a+\frac{R^2}{2}+\frac{R^{\prime 2}}{2})}\right)\right]
\Biggr\}~.
\label{om22}
\end{eqnarray}
\endnumparts
Optical limit of the \eref{opt} corresponds to the
structureless  hadrons  whereas
equations (\ref{chsh}) and (\ref{exp}) imply constituent quark model
.

\section{Inelastic corrections to double scattering}
Partonic structure of the constituent quark is a natural
bridge between constituent quarks in the bound-state
problem of the hadrons and the partons as probed in
deep-inelastic scattering.
For the constituent quarks besides an elastic scattering
the inelastic one (with excitation one or both quarks) exists.
It means that under collision
process
quark can transit to an excited state, that
afterwards leads to the production of new particles. But if the excited
quark strikes once more it can lose excitation and makes
the contribution to the elastic hadron scattering
amplitude. The similar effects were considered for nuclei collision
(Gribov 1969, Karmanov and Kondratyuk 1973)
on a nucleon level, and in the problem of
"color transparency" (Benhar \etal 1996,
Nikolaev 1993).

Formula (\ref{om2}) describes double scattering of three kinds shown at
figure 1. We
should add the diagrams of figure 2. to the diagrams of
figure 1. As a result the \eref{om2} becomes:
\begin{eqnarray}
\fl
(\Omega^\prime_m({\bf b}))^2\equiv
-\frac{18}{4\pi^2}\{\int (f({\bf q})f({\bf q}^\prime)+\sum_{n\ne 0}
f_{0n}({\bf q})f_{n0}({\bf q}^\prime))
[G^\prime ({\bf q+q}^\prime)G_m ({\bf q,q}^\prime)+
\nonumber
\\ \ms
+G^\prime ({\bf q,q}^\prime)G_m ({\bf q+q}^\prime)]
\e^{-\i({\bf q+q}^\prime){\bf b}}\d^2{\bf q}\d^2{\bf q}^\prime
\nonumber
\\ \ms
+
2\int f({\bf q})f({\bf q}^\prime)G_m({\bf q,q}^\prime) G^\prime
({\bf q,q}^\prime)\e^{-\i({\bf q+q}^\prime){\bf b}}\d^2
{\bf q}\d^2{\bf q}^{\prime}\}.
\end{eqnarray}
The amplitude $ f _{n0} (q) ~~ (f _{00} (q) \equiv f (q)) $
represents the quark scattering amplitude
corresponding to the one
quark transition to the n-th
excited state and  $ f _{0n} (q) ~$ is
amplitude of the back transition from the excited state to the unexcited
one.
Let's consider the contribution of this inelastic process to
the elastic forward $ \Omega ^{\hbox{-}}p $ scattering amplitude taking
into account only double collisions:
\begin{eqnarray}
\fl
F_m^{(2)}(0)=\frac{\i}{2\pi}\int \d^2{\bf b}
\frac{1}{2}\Omega_m^{\prime 2}({\bf b})=
\frac{-18\i}{4\pi}\Biggl[ \int\sum_{n=0}f_{0n}({\bf q})
f_{n0}(-{\bf q}) \biggl(
G^{\prime}(0)G_m({\bf q,-q})
\nonumber
\\
+G^{\prime}({\bf q,-q})G_m(0)\biggr) \d^2{\bf q} +
2\int f({\bf q})f({\bf q}^{\prime})G^{\prime}
({\bf q,-q})G_m({\bf q,-q})\d^2{\bf q}\Biggr]
\nonumber
\\
\approx
\frac{-18\i}{4\pi}\Biggl[\sum_{n=0}f_{0n}(0)f_{n0}(0)
\int\biggl( G^{\prime}(0)
G_m({\bf q,-q})
\nonumber
\\
+G^{\prime}({\bf q,-q})G_m(0)\biggr) \d^2{\bf q} +
2f(0)f(0)\int G^{\prime}({\bf q,-q})G_m({\bf q,-q})\d^2{\bf q}
\Biggr]~.
\label{F2in}
\end{eqnarray}
Under derivation of \eref{F2in}
we assume that the form factors of
a proton and
hyperon are characterized by sharper $q-$ dependence in comparison with
the amplitudes $ f _{0n} ({\bf q}) $ taken out of the
integral. So, the inelastic corrections  are mainly determined by
the amplitude of inelastic quark scattering at zero angle.

Let's consider a model which follows from the naive concept of
composite particles, i.e. from assumption, that
the constituent quark consists of elementary constituents (partons).
Partonic picture of hadron collision is not fully invariant under Lorentz
transforms (Gribov 1973). We use the centrer of mass frame.

Among all partons there are $ N$ active partons,
which can elastically scatter each other under the quark collision.
The active partons is considered
to be slow, and carry a small part of hadron
momentum (Gribov 1973). If the quark excitation
energy is about a pion mass  $ m_ {\pi} $, then wave function of
excited state
of quark depends on time as
$ \sim \e ^ {\i m _ {\pi} t}$.
 One oscillation  is made on length 1 $ /m _ {\pi} $ in a
quark rest frame, but in a laboratory frame this length increases
in $ \gamma $ times ($ \gamma $ is
Lorentz - factor of a particle) and becomes much greater than
the hadron size $ 1/m _ {N} $. It allows us to neglect the evolution
of quark inner degrees
freedom during the collision. That is a usual condition of
Glauber approximation applicability.
A peculiarity of the partonic level compared to the
constituent quark level
is the fluctuating number of the active partons in the
constituent quark.
We may write the
zero-angle quark  scattering amplitude as:
\begin{eqnarray}
\hat f(0)=\frac{\i}{2\pi}\int \d^2{\bf b}
\Gamma(\hat N,\hat N^\prime,{\bf b}) ,
\nonumber
\\ \ms
\fl
\Gamma(\hat N,\hat N^\prime,{\bf b})
=-\sum_{k=1}^{\hat N*\hat N^\prime}
\frac{\hat N\hat N^\prime(\hat N\hat N^\prime-1)...
(\hat N\hat N^\prime-k+1)}{k!}
\Biggl(-\frac{3\mbox{f}_p}{2\i r_q^2}
\exp{\left(-\frac{3b^2}{4r_q^2}\right)}\Biggr)^k
\nonumber
\\
=1 -\Biggl(1-\frac{3\mbox{f}_p}{2\i r_q^2}
\exp{\left(-\frac{3b^2}{4r_q^2}\right)}\Biggr)^{\hat N \hat N^{\prime}}
,
\end{eqnarray}
where $r_q\approx\frac{3}{2}a$ is mean square radius of constituent
quark  and $\mbox{f}_p$ is amplitude of the parton
elastic scattering.
The amplitude $ \hat f (0) $
and profile-function $ \Gamma({\bf b}) $
contain operators of the active parton number
$\hat N$ and $\hat N^\prime$
in a hyperon and proton quarks, respectively.
Because $b$-picture of quark collision is not
important here
we simplify it using single-scattering eikonal function in
degree of collision multiplicity $k$ and then multiply it
by the number of the $k$-multiple terms.
For quark elastic scattering
amplitude  we have:
\begin{equation}
f(0)=<0,0\mid \hat f(0)
\mid 0,0>=
\frac{\i}{2\pi}\sum_{N,N^\prime}P(N)P(N^\prime)\int\Gamma(N,N^\prime,b)d^2b.
\label{f00}
\end{equation}
$\mid 0,0>\equiv\mid 0>\mid 0>$ , $\mid 0>$ 
is unexcited state of the constituent
quark,
$P(N)$ is probability to find $N$ active partons in the
constituent quark.
Now let's consider the magnitude of
$ < 0,0\mid\hat f (0)\mid 0>< 0\mid \hat f (0) \mid 0,0> $,
containing the contribution of inelastic corrections, which is
$ \sum _ {n\ne 0}
 < 0,0\mid \hat f (0) \mid n,0>< 0,n\mid \hat f (0)\mid 0,0> $.
The direct evaluation gives:
\begin{eqnarray}
<0,0\mid \hat f(0)\mid 0><0\mid \hat f(0)\mid 0,0>
\nonumber
\\
=
\frac{-1}{4\pi^2}\sum_{N,N^\prime,N^{\prime\prime}}
P(N)P(N^\prime)P(N^{\prime\prime})\int\d^2{\bf b}\d^2{\bf b}^\prime
\Gamma(N,N^\prime,b)\Gamma(N,N^{\prime\prime},b^\prime).
\label{df0}
\end{eqnarray}
By reversing \eref{f00} it is possible
to find $\mbox{f}_p$ through $f(0)$ and substituting it in
\eref{df0} find
$<0,0\mid \hat f(0)\mid 0><0\mid \hat f(0)\mid 0,0>$.
Double quark scattering with inelastic
shadowing can be roughly found by using
$<0,0\mid \hat f(0)\mid 0><0\mid \hat f(0)\mid 0,0> $
in two first items in \eref{om22} instead of $f^2(0)$.
Such a program was realized in our numerical
calculations.

\section{ Results and discussion}
We now substitute $ \Omega ^ {\prime 2} _
{_ {\stackrel {3/2} {1/2}}} ({\bf b}) $
and $ \Omega _ {_ {\stackrel {3/2} {1/2}}} ({\bf b}) $ from
equations
(\ref{om11}), (\ref{om22}) in one of the equations
(\ref{chsh}), (\ref{exp}), (\ref{opt}), (\ref{double})
for $ \gamma _ m
(b) $ and through numerical integration of eref{fmm}  find the
difference between the forward elastic $ \Omega ^ {\hbox {-}}-p $ scattering
amplitudes corresponding to hyperon spin projections $ \pm $ 1/2 and $ \pm $
3/2. Then we are able to obtain the magnitude of dichroism and spin
oscillation phase to be interested in. For a determinacy we shall make the
calculations for the meter hydrogen target of $0.0675~ g/cm ^ {3} $ density.
The value of a quadrupole moment $ Q  = 2  \times 10  ^ {-2} (fm) ^ 2 $
we shall take from work (Gershtein and Zinoviev 1981).  The mean squared
radiuses of a hyperon and proton we shall accept as 0.35 fermi squared
(Gershtein and Zinoviev 1981) and 0.65 (Furget \etal 1990) respectively.
For the constants $ f(0) $ and $ a $ we have from (\ref{ami}):
\begin{equation}
\eqalign{ f(0)=(5.6\times10^{-3}~,~2.8\times10^{-2})~~~~(fm)^2~,
\\ a=3.1\times10^{-2}~~~~(fm)^2.  }
\end{equation}
In the table 1. it is
shown the  difference between the zero angle scattering amplitudes (real
and imaginary parts).  as well as dichroism and phase of spin oscillation of
hyperon with the energy $292.4~Gev$ calculated with the various forms of
function $\gamma_m(b)$.  $ \phi $ is the phase of oscillations of a hyperon
spin:
\begin {equation} \phi (z)=2\pi \rho Re(F_{3/2}(0)-F_{1/2}(0))z,
\label{fi}
\end{equation}
and
\begin{equation}
A(z)={I_{1/2}(z)-I_{3/2}(z)\over I_{3/2}(z)+I_{1/2}(z)} =
2\pi \rho Im(F_{3/2}(0)-F_{1/2}(0))z
\label{dih}
\end{equation}
describes dichroism. $ I _ m (z) $ is the intensity of hyperons with
the
spin projection $ m $ on the depth $ z $ , if
the incident beam is unpolarized.
Let us discuss the dependence of the effect i.e.
dependence of the amplitude difference  $F_{3/2}(0)-F_{1/2}(0)$
on the hyperon energy.
In accordance with the
Regge (Collins and Squires 1968) and parton (Gribov 1973)
theories the quantity $a$, proportional to the
sum of mean squared radiuses of colliding quarks, grows as $ln(s)$,
so, we can write roughly:
\begin{equation}
a(s^q)=a(s^q_0)\left( 1+ln\left( \frac{s^q}{s^q_0} \right) \right).
\end{equation}
$\sqrt{s ^ q} $ is the energy of colliding quarks in a system of their
center-of-mass, $a $ is the known magnitude of
$ a (s ^ q) $, for a given $ s ^ q_ 0 $
.
At high energies $ \frac {s ^ q} {s ^ q _ 0} =
\frac {s} {s _ 0} $, where $ s $ is the squared energy  of
colliding hadrons
 in a center-of-mass frame. We take $ \sqrt s _ 0 = 23.5~ Gev $)
As for the quarks scattering cross section
 $ \sigma _ {qs} $, used in \eref{ami}, it is possible to write
from the
Regge theory :
\begin {equation}
\sigma _ {qs} (s ^ q) = \sigma _ {qs} (s ^ q _ 0)
\left (\frac {s ^ q} {s ^ q _
0} \right) ^ \Delta. \end {equation}
We have taken the value $ \Delta = 0.125 $ from  $ p-p $ scattering
data (Gotsman \etal 1994).
The ratio of the real part of quark scattering
amplitude to the imaginary one is
considered to be value as
at $ \sqrt {s} = 23.5~ Gev $ i.e. 0.2.
The dependence of phase oscillation (it is approximately
equaled to a rotation angle) and dichroism on hyperon energy
are
shown on figures 3,4 respectively.
Curve (a) correspounds the
optical limit, i.e. structureless hadrons and
lies considerably lower then the curves (b),(c) suggesting
constituent quark model.  
The inelastic corrections (curve (B), (C)) evaluated
for Poisson distribution for the number of active
partons in quarks with mean active parton number equal 1.
If mean number of active partons
grow the inelastic corrections decreases.
The difference between (b) and (c) as well as (B) and (C) arises due to
different manner of three- and more- multiple
collision account.
Let's return to spin dynamics of $ \Omega ^ {\hbox {-}} $-hyperon.
If the initial polarization vector of a hyperon lays
on the plane $ x, z $
and is directed at an angle $ \theta $ to the axes $ z $,
the hyperon initial spin wave function is determined as:
\[
\hat\psi _ 0 = \{ d ^ {3/2} _ {3/2 ~ 3/2} (\theta), d ^ {3/2} _ {1/2~3/2 }
(\theta), d ^ {3/2} _ {-1/2~3/2} (\theta), d ^ {3/2} _ {-3/2~3/2} (\theta) \},
\]
where $ d ^ S _ {M \, M ^ \prime} (\theta) $ is
d-function of the finite rotation matrix.
(Varshalovich \etal 1975).
 Using equation (6) we obtain:
\begin{equation}
\eqalign{
\fl ~~~~~~~~~~~
<S_x>=\frac{3}{4}\{\sin^3{\theta}\,\e^{-2 n^{\prime\prime}_{1/2} kz}+
\sin{\theta}(1+\cos{\theta})
\,\e^{- (n^{\prime\prime}_{3/2} +  n^{\prime\prime}_{1/2}) kz}
\cos{\phi}\}/|\psi|^2,
\\ \ms
\fl ~~~~~~~~~~~
<S_y>=-\frac{3}{2}\sin{\theta}\cos{\theta}
\,\e^{-  (n^{\prime\prime}_{3/2} +  n^{\prime\prime}_{1/2})kz}
\sin{\phi}/|\psi|^2,
\\ \ms
\fl ~~~~~~~~~~~
<S_z>=\{\frac{3}{4}\cos{\theta}(1+\cos^2{\theta})
\,\e^{-2\, n^{\prime\prime}_{3/2}\,kz}
+\frac{3}{8}\sin^2{\theta}\cos{\theta}
\,\e^{-2\, n^{\prime\prime}_{1/2}\,kz}\}/|\psi|^2;
\\ \ms
|\psi|^2=(1-\frac{3}{4}\sin^2{\theta})
\,\e^{-2\, n^{\prime\prime}_{3/2}\,kz} +
\frac{3}{4}\sin^2{\theta}
\, \e^{-2\, n^{\prime\prime}_{1/2}\,kz}.
}
\label{spi}
\end{equation}
From equations (\ref{spi}) we see that
under the acute $ \theta $ angle, initially zero quantity
$ < S _ y > $ receives a negative increment,
and under obtuse $ \theta $ - a positive increment when
$\phi$ increases.
So in the first case the spin rotates
counter-clockwise, and in second - clockwise if to look against
hyperon motion.
Initially zero quantity
$ < Q _ {yz} > $ receives a negative
increment in the first case and positive in second.
Increment of initially zero magnitude $ < Q _ {xy} > $ is negative
in the both cases. At $ \theta = \pi/2 $
$ < S _ y > $, $ < S _ z> $, $ < Q _ {xy} > $, $ < Q _ {xz} > $
remains zero. Initially zero magnitude of
$ < Q _ {yz} > $ receives an increment. The remaining
quantities vary from nonzero initial values.

For a hyperon with energy $292.4~ Gev~ (\sqrt {s} = 23.5 ~ Gev )$
in a meter hydrogen target
dichroism
$ \approx 1.4\times 10 ^ {-4} ~$
for the structureless hadrons and
$  2.0-2.1\times 10 ^ {-4} ~$ for the  hadrons
built from the constituent quarks.
Oscillation phase (rotation angle) is
$ \approx 5.2\times 10 ^ {-5} $ in the
first case and $ \approx 7.1-7.7 \times 10 ^ {-5} ~$
in the second.
There is good reason to believe
that they will be greater due to
inelastic corrections.
We may use any light nucleus target
instead of hydrogen one. For light
nuclei forward scattering amplitude
is proportional to the mass number,
so  the value of rotation angle
and dichroism may be obtained by using
nucleon density in the target chosen as $\rho$
in equations (\ref{fi}), (\ref{dih}).
In a meter carbon target of $2.2~g/cm^3$ density dichroism
$ \approx 7.5\times 10 ^ {-3}~ $
and rotation angle $ \approx 2.5\times 10 ^ {-3}~ $.
For measuring rotation angle of the
$ \Omega ^ {\hbox {-}} $-hyperon polarization
 procedure used in (Diehl \etal 1991) is
applicable although higher statistics is needed.

\section{ Conclusion }
Summing up it is possible to tell that
the spin oscillations and dichroism,
of high-energy particles with a spin $S \geq 1 $ passing
through matter allow  one to
observe constituent
quark rescattering directly in a broad
energy range 
avoiding a background of
single scattering.  An importan point is that the effect increases
the energy increases.
\section*{Appendix}
The $3/2$ spins  matrixes look like:
\begin{eqnarray}
\hat S_x=\left(
\begin{array}{cccc}
~~0~~ & {{{\sqrt{3}}}\over 2} & ~~0~~ & ~~0~~ \\
{{{\sqrt{3}}}\over 2} & 0 & 1 & 0 \\
0 & 1 & 0 & {{{\sqrt{3}}}\over 2} \\
0 & 0 & {{{\sqrt{3}}}\over 2} & 0
\end{array}\right)
~~~
\hat S_y=\left(
\begin{array}{cccc}
0 &{{-\i}\over 2}{\sqrt{3}} & 0 & 0 \\
{\i\over 2}{\sqrt{3}} & 0 &-\i & 0  \\
0 &\i & 0 & {{-\i}\over 2}{\sqrt{3}} \\
0 & 0 &{\i\over 2}{\sqrt{3}} &0
\end{array}\right)
\nonumber
\\ \bs
\hat S_z=\left(
\begin{array}{cccc}
{3\over 2}&0&0&0\\
~~0~~ & {1\over 2} & ~~0~~ & ~~0~~ \\
0 & 0 &-{1\over 2} & 0 \\
0 & 0 & 0 & -{3\over 2}
\end{array}\right)~.
\nonumber
\end{eqnarray}
For the full description
of spin properties of $ \Omega^{\hbox{-}} $-hyperon
as a particle of a spin 3/2 the knowledge of the mean values of
quadrupole tensor
\[
\hat Q _ {ij} =\frac{1}{2}( \hat S _ i \hat S _ j + \hat S _ j \hat S _ i)
- \frac{5}{4} \delta _ {ij}
\]
and octupole tensor
\[
\hat T_{ijk}=
\frac{5}{27} \left(
 Perm \{\hat S_i \hat S_j \hat S_k\} -\frac{41}{10}
(\delta_{ij}\hat S_{k} + \delta_{jk}\hat S_{i} + \delta_{ki}\hat S_{j})
\right)
\]
components
 are necessary.
Here $ Perm $ means all possible permutations of indexes.
Averaging these operators over wave function (2), we obtain
(without coping
the values of polarization components already listed in (6) ):

\begin{eqnarray}
\fl
<\hat Q_{xy}>=-\sqrt{3}\{ \mbox{a}\mbox{c}
   \sin (\delta_{3/2} - \delta_{-1/2} + (n^{\prime}_{3/2} -
     n^{\prime}_{1/2}) z )
\nonumber
\\ \bs
+ \mbox{b}\mbox{d}\sin
(\delta_{1/2} - \delta_{-3/2} -(n^{\prime}_{3/2}-n^{\prime}_{1/2})kz)
 \} \e^{-(n^{\prime \prime}_{3/2} +n^{\prime \prime}_{1/2})kz}
/|\psi^2|,
\nonumber
\\ \bs
\fl
<\hat Q_{xz}>=\sqrt{3}\{ \mbox{a}\mbox{b}
   \cos ( \delta_{3/2} - \delta_{1/2} + (n^{\prime}_{3/2} -
      n^{\prime}_{1/2})kz)
\nonumber
\\ \bs
-\mbox{c}\mbox{d}
\cos ( \delta_{-1/2} - \delta_{-3/2} - (n^{\prime}_{3/2}-
      n^{\prime}_{1/2})kz)\} \e^{-(n^{\prime \prime}_{3/2}
+n^{\prime \prime}_{1/2})kz} / |\psi^2|,
\nonumber
\end{eqnarray}
\begin{eqnarray}
\fl
<\hat Q_{yz}>=\sqrt{3}\{ -\mbox{a}\mbox{b}
   \sin ( \delta_{3/2} -  \delta_{1/2} + ( n^{\prime}_{3/2} -
      n^{\prime}_{1/2})kz)
\nonumber
\\ \bs
+\mbox{c}\mbox{d}
\sin ( \delta_{-1/2} - \delta_{-3/2} - (n^{\prime}_{3/2}-
      n^{\prime}_{1/2})kz )
\e^{-(n^{\prime \prime}_{3/2}
+n^{\prime \prime}_{1/2})kz}\} / |\psi^2|,
\nonumber
\\ \bs
\fl
<\hat Q_{xx}>=\frac{1}{2}
\{
(-\mbox{a}^2 -
  \mbox{d}^2) \e^{-2 n^{\prime \prime}_{3/2}kz} +
 ( \mbox{b}^2 +
  \mbox{c}^2) \e^{-2 n^{\prime \prime}_{1/2}kz}
\nonumber
\\ \bs
+ 2 \sqrt{3} [\mbox{a}\mbox{c}
   \cos ( \delta_{3/2} -  \delta_{-1/2} + (n^{\prime}_{3/2} -
      n^{\prime}_{1/2})kz)
\nonumber
\\ \bs
+ \mbox{b}\mbox{d}
   \cos (\delta_{1/2} -\delta_{-3/2} - (n^{\prime}_{3/2}-
      n^{\prime}_{1/2})kz)]\e^{-(n^{\prime \prime}_{3/2}+
n^{\prime \prime}_{1/2})kz}
\} / |\psi^2|,
\nonumber
\\ \bs
\fl
<\hat Q_{yy}>=
\frac{1}{2}
\{
(-\mbox{a}^2 -
  \mbox{d}^2) \e^{-2 n^{\prime \prime}_{3/2}kz} +
 ( \mbox{b}^2 +
  \mbox{c}^2) \e^{-2 n^{\prime \prime}_{1/2}kz}
\nonumber
\\ \bs
- 2 \sqrt{3}[\mbox{a}\mbox{c}
   \cos ( \delta_{3/2} -  \delta_{-1/2} + (n^{\prime}_{3/2} -
      n^{\prime}_{1/2})kz)
\nonumber
\\ \bs
+ \mbox{b}\mbox{d}
   \cos (\delta_{1/2} -\delta_{-3/2} - (n^{\prime}_{3/2}-
      n^{\prime}_{1/2})kz)] \e^{-(n^{\prime \prime}_{3/2}+
 n^{\prime \prime}_{1/2})kz}
\} / |\psi^2|,
\nonumber
\end{eqnarray}
\begin{eqnarray}
\fl
<\hat Q_{zz}>=
\{
(\mbox{a}^2 +
  \mbox{d}^2)\e^{-2 n^{\prime \prime}_{3/2}kz} -
  (\mbox{b}^2 +
  \mbox{c}^2)\e^{-2 n^{\prime \prime}_{1/2}kz}
\}/ |\psi^2|,
\nonumber
\\ \bs
\fl
<\hat T_{xyz}>=\frac{5}{3\sqrt{3}}
\{-
\mbox{a}\mbox{c}
\sin (\delta_{3/2} -  \delta_{-1/2} +
(n^{\prime}_{3/2} -  n^{\prime}_{1/2})kz)
\nonumber
\\ \bs
 + \mbox{b}\mbox{d}
   \sin ( \delta_{1/2} -  \delta_{-3/2} - (n^{\prime}_{3/2}-
      n^{\prime}_{1/2})kz)
\}\e^{-(n^{\prime \prime}_{3/2}
+n^{\prime \prime}_{1/2})kz} / |\psi^2|,
\nonumber
\\ \bs
\fl
<\hat T_{xxy}>=
\{
-\frac{1}{3} \e^{-2 n^{\prime \prime}_{1/2}kz}
\mbox{b}\mbox{c} \sin ( \delta_{1/2} -  \delta_{-1/2})
-\frac{5}{3}\e^{-2n^{\prime \prime}_{3/2}kz}
\mbox{a}\mbox{d} \sin ( \delta_{3/2} -  \delta_{-3/2})
\nonumber
\\ \bs +         \frac{1}{3\sqrt{3}}
 [\mbox{a}\mbox{b}
      \sin ( \delta_{3/2} -  \delta_{1/2} +(n^{\prime}_{3/2} -
         n^{\prime}_{1/2})kz)
\nonumber
\\ \bs +
  \mbox{c}\mbox{d}
\sin \{ \delta_{-1/2} - \delta_{-3/2} -(n^{\prime}_{3/2}-
         n^{\prime}_{1/2})kz)]\e^{-(n^{\prime \prime}_{3/2}+
           n^{\prime \prime}_{1/2})kz}
\}/ |\psi^2|,
\nonumber
\end{eqnarray}
\begin{eqnarray}
\fl
<\hat T_{yyy}>=
\{
- \e^{-2 n^{\prime \prime}_{1/2}kz}
\mbox{b} \mbox{c} \sin ( \delta_{1/2} -  \delta_{-1/2})
+ \frac{5}{3}\e^{-2n^{\prime \prime}_{3/2}kz}
\mbox{a}\mbox{d} \sin ( \delta_{3/2} -  \delta_{-3/2})
\nonumber
\\ \bs
+         \frac{1}{\sqrt{3}}
 [\mbox{a}\mbox{b}
      \sin ( \delta_{3/2} -  \delta_{1/2} +(n^{\prime}_{3/2} -
         n^{\prime}_{1/2})kz)
\nonumber
\\ \bs
+  \mbox{c}\mbox{d}
\sin ( \delta_{-1/2} - \delta_{-3/2} -(n^{\prime}_{3/2}-
         n^{\prime}_{1/2})kz)]\e^{-(n^{\prime \prime}_{3/2}+
           n^{\prime \prime}_{1/2})kz}
\} / |\psi^2|,
\nonumber
\end{eqnarray}
\begin{eqnarray}
\fl
<\hat T_{zzy}>=
\{
\frac{4}{3}\e^{-2 n^{\prime \prime}_{1/2}kz}
\mbox{b}\mbox{c} \sin ( \delta_{1/2} -  \delta_{-1/2})
\nonumber
\\ \bs
-          \frac{4}{3\sqrt{3}}
 [\mbox{a}\mbox{b}
      \sin ( \delta_{3/2} -  \delta_{1/2} +(n^{\prime}_{3/2} -
         n^{\prime}_{1/2})kz)
\nonumber
\\ \bs
+  \mbox{c}\mbox{d}
\sin ( \delta_{-1/2} - \delta_{-3/2} -(n^{\prime}_{3/2}-
         n^{\prime}_{1/2})kz)] \e^{-(n^{\prime \prime}_{3/2}+
           n^{\prime \prime}_{1/2})kz}
\}/ |\psi^2|,
\nonumber
\\ \bs
\fl
<\hat T_{xxx}>=
\{
\mbox{b}\mbox{c}\e^{-2 n^{\prime \prime}_{1/2}kz}
\cos ( \delta_{1/2} -  \delta_{-1/2}) +
  \frac{5}{3}\mbox{a}\mbox{d}\e^{-2 n^{\prime \prime}_{3/2}kz}
\cos ( \delta_{3/2} -  \delta_{-3/2})
\nonumber
\\ \bs
  -
  \frac{1}{ \sqrt{3}}[\mbox{a}\mbox{b}
      \cos ( \delta_{3/2} -  \delta_{1/2}+ (n^{\prime}_{3/2} -
         n^{\prime}_{1/2})kz)
\nonumber
\\ \bs +
  \mbox{c}\mbox{d}
      \cos (\delta_{-1/2} - \delta_{-3/2} - (n^{\prime}_{3/2}-
         n^{\prime}_{1/2})kz)] e^{-(n^{\prime \prime}_{3/2}+
           n^{\prime \prime}_{1/2})kz}
\}/ |\psi^2|,
\nonumber
\\ \bs
\fl
<\hat T_{zzz}>=
\{
\frac{1}{3}(\mbox{a}^2 -
  \mbox{d}^2)\e^{-2n^{\prime \prime}_{3/2}kz} +
(-\mbox{b}^2 +
 \mbox{c}^2)\e^{-2n^{\prime \prime}_{1/2}kz}
\}/|\psi|^2~,
\nonumber
\\ \bs
\fl
<\hat T_{zzx}>=
\{
-\frac{4}{3}\mbox{b}\mbox{c}\e^{-2 n^{\prime \prime}_{1/2}kz}
\cos (\delta_{1/2} -
 \delta_{-1/2})
\nonumber
\\ \bs
+
\frac{4}{3\sqrt{3}}
[\mbox{a}\mbox{b}
      \cos (\delta_{3/2} - \delta_{1/2} +(n^{\prime}_{3/2} -
        n^{\prime}_{1/2})kz)
\nonumber
\\ \bs +
  \mbox{c}\mbox{d}
\cos ( \delta_{-1/2} - \delta_{-3/2} - (n^{\prime}_{3/2}-
         n^{\prime}_{1/2})kz)]\e^{-(n^{\prime \prime}_{3/2}+
           n^{\prime \prime}_{1/2})kz}
\}/|\psi|^2~,
\nonumber
\\ \bs
\fl
<\hat T_{xxz}>=
\{
\frac{1}{6}(-\mbox{a}^2 +
  \mbox{d}^2)\e^{-2n^{\prime \prime}_{3/2}kz} +
  \frac{1}{2}(\mbox{b}^2 -
  \mbox{c}^2)\e^{-2n^{\prime \prime}_{1/2}kz}
\nonumber
\\ \bs  +
  \frac{5}{3\sqrt{3}}[\mbox{a}\mbox{c}
   \cos (\delta_{3/2} - \delta_{-1/2} + (n^{\prime}_{3/2} -
      n^{\prime}_{1/2})kz)
\nonumber
\\ \bs -
\mbox{b}\mbox{d}
   \cos (\delta_{1/2} - \delta_{-3/2} - (n^{\prime}_{3/2}-
     n^{\prime}_{1/2})kz)]
\e^{-(n^{\prime \prime}_{3/2}+
           n^{\prime \prime}_{1/2})kz}
\}/|\psi|^2~,
\nonumber
\\ \bs
\fl
<\hat T_{yyz}>=
\{
\frac{1}{6}(-\mbox{a}^2 +
  \mbox{d}^2)\e^{-2n^{\prime \prime}_{3/2}kz} +
  \frac{1}{2}(\mbox{b}^2 -
  \mbox{c}^2)\e^{-2n^{\prime \prime}_{1/2}kz}
\nonumber
\\ \bs
+
  \frac{5}{3\sqrt{3}}[-\mbox{a}\mbox{c}
   \cos (\delta_{3/2} - \delta_{-1/2} + (n^{\prime}_{3/2} -
      n^{\prime}_{1/2})kz)
\nonumber
\\ \bs
+
\mbox{b}\mbox{d}
   \cos (\delta_{1/2} - \delta_{-3/2} - (n^{\prime}_{3/2}-
     n^{\prime}_{1/2})kz)]
\e^{-(n^{\prime \prime}_{3/2}+
           n^{\prime \prime}_{1/2})kz}
\}/|\psi|^2~.
\nonumber
\end{eqnarray}

\References
\item[] Yndurain F J 1993
     {\it The theory of quark and gluon interaction}
     (Berlin: Springer-Verlag)
\item[] Akchurin N \etal 1993 {\it Phys. Rev. D} {\bf 48} 3026
\item[] Baryshevsky V G 1992 {\it Phys.Lett.} {\bf 171A} 431
\item[] \dash 1993 {\it J.Phys.G} {\bf 19} 273
\item[] Anisovich V V \etal 1985
{\it Quark Model and High Energy Collisions}
(Singapore: World Scientific)
\item[] Gell-Mann M 1964 {\it Phys. Lett.} {\bf 8} 214
\item[] Hwa R C 1980  {\it Phys. Rev. D} {\bf 22} 1593
\item[] Czyz W and Maximon L C 1969 {\it Ann.Phys.} {\bf 52} 59
\item[] Lax M 1951  {\it Rev. Mod. Phys.} {\bf 23} 287
\item[] Furget C, Buenerd M and Valin P 1990 {\it Z.Phys.C} {\bf 47} 377
\item[] Gershtein S S and Zinoviev G M 1981
{\it Sov. J. Nucl. Phys.} {\bf 33} 772 ({\it Yad.Fiz.} {\bf 33} 1442)
\item[] Franco V and Varma G K 1978 {\it Phys. Rev. C} {\bf 18} 167
\item[] Gribov V N 1969 {\it JETP} {\bf 29} 483
        ({\it Zh. Eksp. Teor. Fiz.} {\bf 56} 892)
\item[] Karmanov V A and Kondratyuk L A 1973
{\it JETP Lett.} {\bf 18} 266
({\it Pis'ma Zh. Eksp. Teor. Fiz.} {\bf 18} 451)
\item[] Benhar O \etal 1996 {\it Zh. Eksp. Teor. Fiz.} {\bf 110} 1933
\item[] Nikolaev N N 1993 {\it Proc. 27th Winter
School of PINP (St-Peterburg)} {\bf 2} 175
\item[] Gribov V N 1973 {\it Proc. 8th Winter
School of LINP (Leningrad)} {\bf 2} 5 (in Russian)
\item[] Collins P D B and Squires E J 1968
{\it Regge poles in particle physics}
(Berlin:Springer-Verlang)
\item[]
Gotsman E, Levin E M and Maor U 1994
         {\it Phys. Rev. D} {\bf 49} 4321
\item[] Varshalovich D A, Moskalev A N and
   Khersonsky V K 1975 {\it Quantum theory of angular momentum}
 (Moscow:Nauka)(in Russian)
\item[] Diehl H T \etal 1991 {\it Phys. Rev. Lett.} {\bf 67} 804
\endrefs

\newcommand{\Lower}[1]{\smash{\lower 1.5ex \hbox{#1}}}
\begin{table}
   \caption{ Amplitude
   difference ($fm^2$) with spin projections
   $3/2$ and $1/2$, oscillation phase (wich is approximately equal
rotation angle) and dichroism
   for meter hydrogen target of $0.0675~ g/cm ^
{3} $ density. All quantities were calculated for the
 different form of
   $\gamma_m (b)$ : column (a) corresponds to the
   equation (23),
   (b) - (22), (c) -(20),
   (d) -(24).
 }
\begin{indented}
\item[]
\begin{tabular}{r r r r r}
\br
$    $ & $ a~ $ & $ b~ $ & $ c~ $ & $ d~ $ \rule{0in}{3ex} \\[1ex] \mr
\Lower{$F_{3/2}(0)-F_{1/2}(0)$}
&$2.0\times 10^{-4}$
&$2.8\times 10^{-4}$&$3.0\times 10^{-4}$&$4.3\times 10^{-4}$ \\
&$5.8\times 10^{-4}$&
$7.9\times 10^{-4}$&$8.2\times 10^{-4}$ &$1.0\times 10^{-3}$\\ \mr
\Lower{$ \phi ~~~~~~~~~~~  $} &
\Lower{$5.2\times 10^{-5}$}&
\Lower{$7.1\times 10^{-5}$}&
\Lower{$7.7\times 10^{-5}$}&
\Lower{$1.1\times 10^{-4}$}\\
\Lower{$ A~~~~~~~~~~~~  $}&
\Lower{$1.4\times 10^{-4}$}&
\Lower{$2.0\times 10^{-4}$}&
\Lower{$2.1\times 10^{-4}$}&
\Lower{$2.6\times 10^{-4}$}\\[2ex] \br
\end{tabular}
\end{indented}
\end{table}

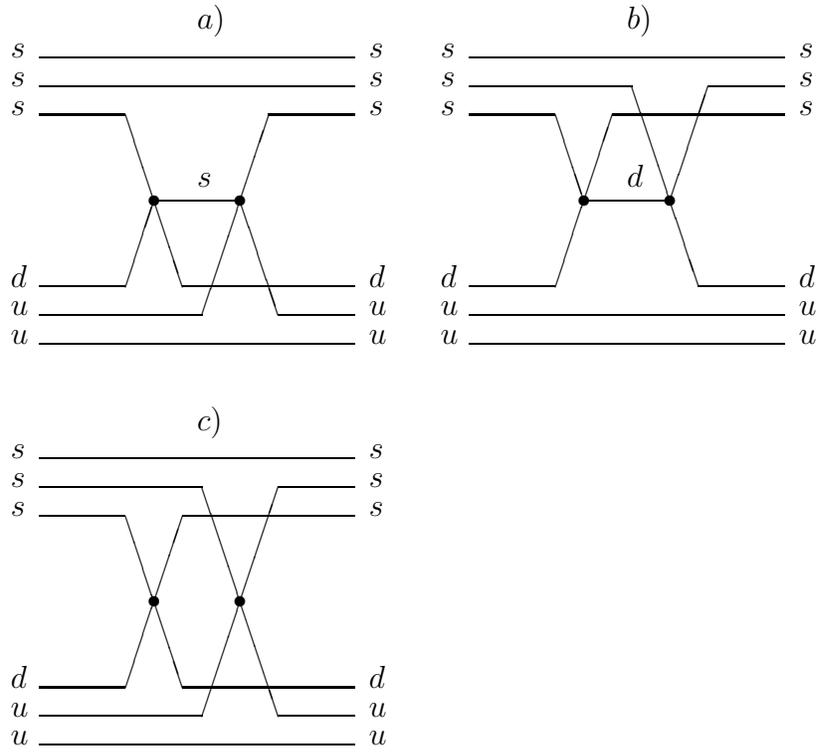
\begin{figure}
\setlength{\unitlength}{0.15in}
\begin{picture}(33,33)(0,0)
\put(2,16){\line(1,0){11}}
\put(2,17  ){\line(1,0   ){5.67  }}
\put(2,18  ){\line(1,0   ){3  }}
\put(2,24   ){\line(1,0   ){3  }}
\put(2,25   ){\line(1,0   ){11  }}
\put(2,26   ){\line(1,0   ){ 11 }}
\put(5,18  ){\line(1,3   ){1  }}
\put(5,24   ){\line(1,-3   ){2  }}
\put(6,21  ){\line(1,0   ){3  }}
\put(7.67,17   ){\line(1,3   ){2.33  }}
\put(9,21  ){\line(1,-3   ){1.33  }}
\put( 7,18  ){\line(1,0   ){6  }}
\put( 10,24  ){\line(1,0   ){ 3 }}
\put(10.33,17   ){\line(1,0   ){2.67  }}
\put(1,24 ){$s$}
\put(1,25 ){$s$}
\put(1,26 ){$s$}
\put(13.5,24 ){$s$}
\put(13.5,25 ){$s$}
\put(13.5,26 ){$s$}
\put(1,16 ){$u$}
\put(1,17 ){$u$}
\put(1,18 ){$d$}
\put(13.5,16 ){$u$}
\put(13.5,17 ){$u$}
\put(13.5,18 ){$d$}
\put(7.5,21.5){$s$}
\put(7.5,27){$a)$}
\put(7.5,13){$c)$}
\put(6,21){\circle*{0.4}}
\put(9,21){\circle*{0.4}}
\put(17,16){\line(1,0){11}}
\put(17,17  ){\line(1,0   ){11  }}
\put(17,18  ){\line(1,0   ){3  }}
\put(17,24   ){\line(1,0   ){3  }}
\put(17,25   ){\line(1,0   ){5.67  }}
\put(17,26   ){\line(1,0   ){ 11 }}
\put(20,18  ){\line(1,3   ){2  }}
\put(20,24   ){\line(1,-3   ){1  }}
\put(21,21   ){\line(1,0   ){3  }}
\put(24,21  ){\line(1,3   ){1.33  }}
\put(22.67,25   ){\line(1,-3   ){2.33  }}
\put( 25,18  ){\line(1,0   ){3 }}
\put( 22,24  ){\line(1,0   ){ 6 }}
\put(25.33,25  ){\line(1,0   ){2.67  }}
\put(16,24 ){$s$}
\put(16,25 ){$s$}
\put(16,26 ){$s$}
\put(28.5,24 ){$s$}
\put(28.5,25 ){$s$}
\put(28.5,26 ){$s$}
\put(16,16 ){$u$}
\put(16,17 ){$u$}
\put(16,18 ){$d$}
\put(28.5,16 ){$u$}
\put(28.5,17 ){$u$}
\put(28.5,18 ){$d$}
\put(22.5,21.5){$d$}
\put(22.5,27){$b)$}
\put(21,21){\circle*{0.4}}
\put(24,21){\circle*{0.4}}
\put(2,2){\line(1,0){11}}
\put(2,3  ){\line(1,0   ){5.67  }}
\put(2,4  ){\line(1,0   ){3  }}
\put(2,10   ){\line(1,0   ){3  }}
\put(2,11   ){\line(1,0   ){5.67  }}
\put(10.33,11){\line(1,0   ){2.67}}
\put(2,12   ){\line(1,0   ){ 11 }}
\put(5,4  ){\line(1,3   ){2  }}
\put(5,10   ){\line(1,-3   ){2  }}
\put(7.67,3   ){\line(1,3   ){2.67  }}
\put(7.67,11   ){\line(1,-3   ){2.67  }}
\put( 7,4 ){\line(1,0   ){6  }}
\put( 7,10  ){\line(1,0   ){ 6 }}
\put(10.33,3   ){\line(1,0   ){2.67  }}
\put(1,10 ){$s$}
\put(1,11 ){$s$}
\put(1,12 ){$s$}
\put(13.5,10 ){$s$}
\put(13.5,11 ){$s$}
\put(13.5,12 ){$s$}
\put(1,2 ){$u$}
\put(1,3 ){$u$}
\put(1,4 ){$d$}
\put(13.5,2 ){$u$}
\put(13.5,3 ){$u$}
\put(13.5,4 ){$d$}
\put(6,7){\circle*{0.4}}
\put(9,7){\circle*{0.4}}
\end{picture}
\caption{Diagrams of the double quark scattering
contribution to the $\Omega^{\hbox{-}}-p$
elastic scattering.}
\end{figure}
\begin{figure}
\setlength{\unitlength}{0.15in}
\begin{picture}(33,33)(0,14)
\put(2, 16){\line(1,0){11}}
\put(2,17  ){\line(1,0   ){5.67  }}
\put(2,18  ){\line(1,0   ){3  }}
\put(2,24   ){\line(1,0   ){3  }}
\put(2,25   ){\line(1,0   ){11  }}
\put(2,26   ){\line(1,0   ){ 11 }}
\put(5,18  ){\line(1,3   ){1  }}
\put(5,24   ){\line(1,-3   ){2  }}
\put(6,21  ){\line(1,0   ){3  }}
\put(7.67,17   ){\line(1,3   ){2.33  }}
\put(9,21  ){\line(1,-3   ){1.33  }}
\put( 7,18  ){\line(1,0   ){6  }}
\put( 10,24  ){\line(1,0   ){ 3 }}
\put(10.33,17   ){\line(1,0   ){2.67  }}
\put(1,24 ){$s$}
\put(1,25 ){$s$}
\put(1,26 ){$s$}
\put(13.5,24 ){$s$}
\put(13.5,25 ){$s$}
\put(13.5,26 ){$s$}
\put(1,16 ){$u$}
\put(1,17 ){$u$}
\put(1,18 ){$d$}
\put(13.5,16 ){$u$}
\put(13.5,17 ){$u$}
\put(13.5,18 ){$d$}
\put(7.5,21.5){$s^{*}$}
\put(7.5,27){$a)$}
\put(6,21){\circle*{0.4}}
\put(9,21){\circle*{0.4}}
\put(6,20.7){\rule{.45in}{0.08in}}
\put(17,16){\line(1,0){11}}
\put(17,17  ){\line(1,0   ){11  }}
\put(17,18  ){\line(1,0   ){3  }}
\put(17,24   ){\line(1,0   ){3  }}
\put(17,25   ){\line(1,0   ){5.67  }}
\put(17,26   ){\line(1,0   ){ 11 }}
\put(20,18  ){\line(1,3   ){2  }}
\put(20,24   ){\line(1,-3   ){1  }}
\put(21,21   ){\line(1,0   ){3  }}
\put(24,21  ){\line(1,3   ){1.33  }}
\put(22.67,25   ){\line(1,-3   ){2.33  }}
\put( 25,18  ){\line(1,0   ){3 }}
\put( 22,24  ){\line(1,0   ){ 6 }}
\put(25.33,25  ){\line(1,0   ){2.67  }}
\put(16,24 ){$s$}
\put(16,25 ){$s$}
\put(16,26 ){$s$}
\put(28.5,24 ){$s$}
\put(28.5,25 ){$s$}
\put(28.5,26 ){$s$}
\put(16,16 ){$u$}
\put(16,17 ){$u$}
\put(16,18 ){$d$}
\put(28.5,16 ){$u$}
\put(28.5,17 ){$u$}
\put(28.5,18 ){$d$}
\put(22.5,21.5){$d^{*}$}
\put(22.5,27){$b)$}
\put(21,21){\circle*{0.4}}
\put(24,21){\circle*{0.4}}
\put(21,20.7){\rule{.45in}{0.08in}}
\end{picture}
\caption{Diagrams of the inelastic corrections to the
double quark scattering.
$ ^*~\mbox{denotes}$ excited quark state. }
\end{figure}
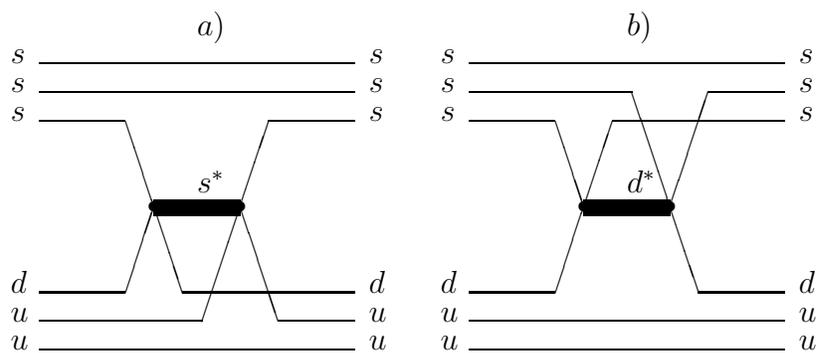
\begin{figure}
\centerline{\epsfbox{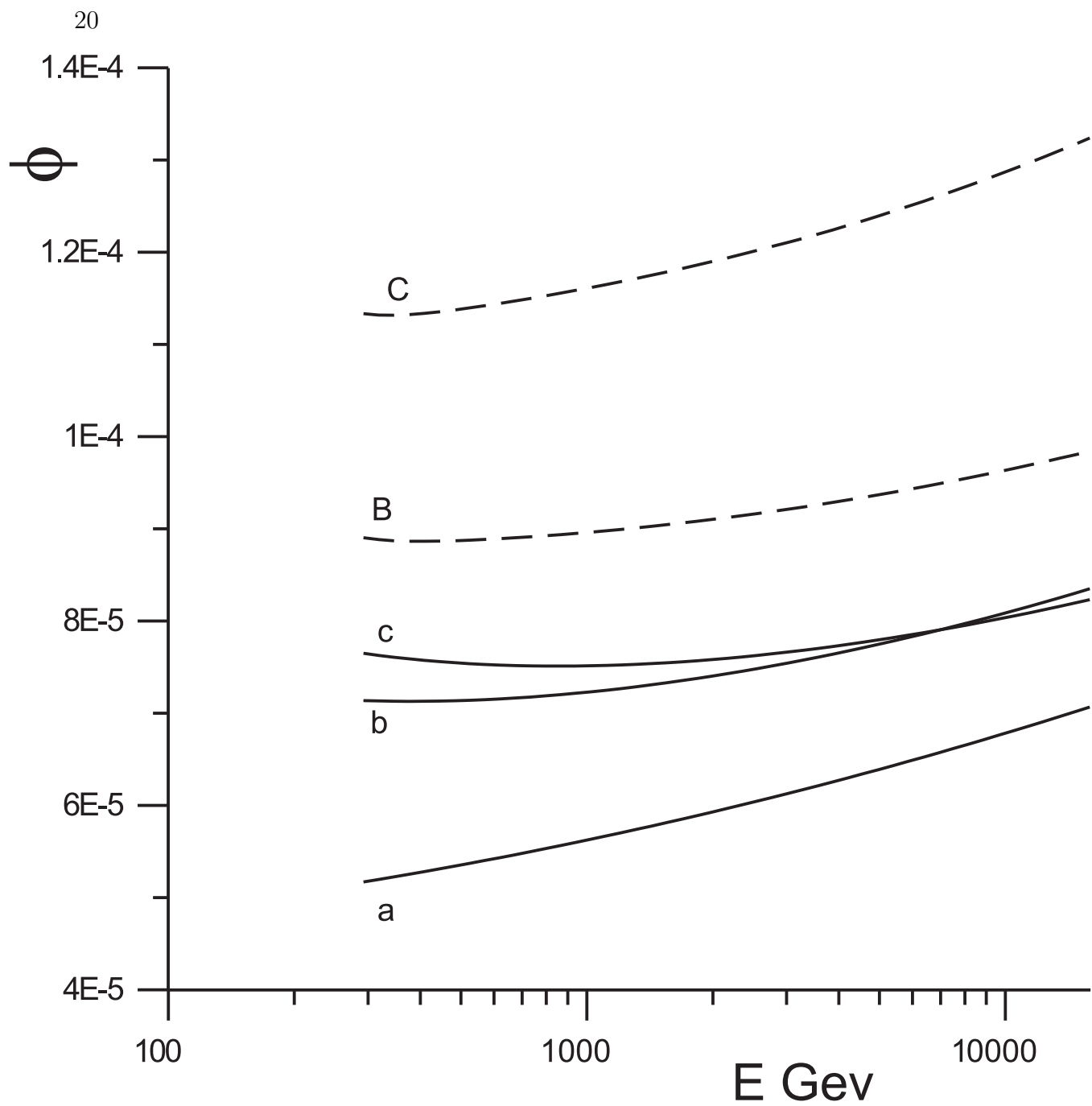}}
\caption{Energy dependence of $\Omega^{\hbox{-}}$-hyperon
spin oscillation phase
(rotation angle) in
hydrogen target, calculated  for the
different form of  $\gamma_m (b)$ :
curve (a) corresponds to the
 equation (23),
   (b,B) - (22), (c,C) -(20).}
\end{figure}
\begin{figure}
\centerline{\epsfbox{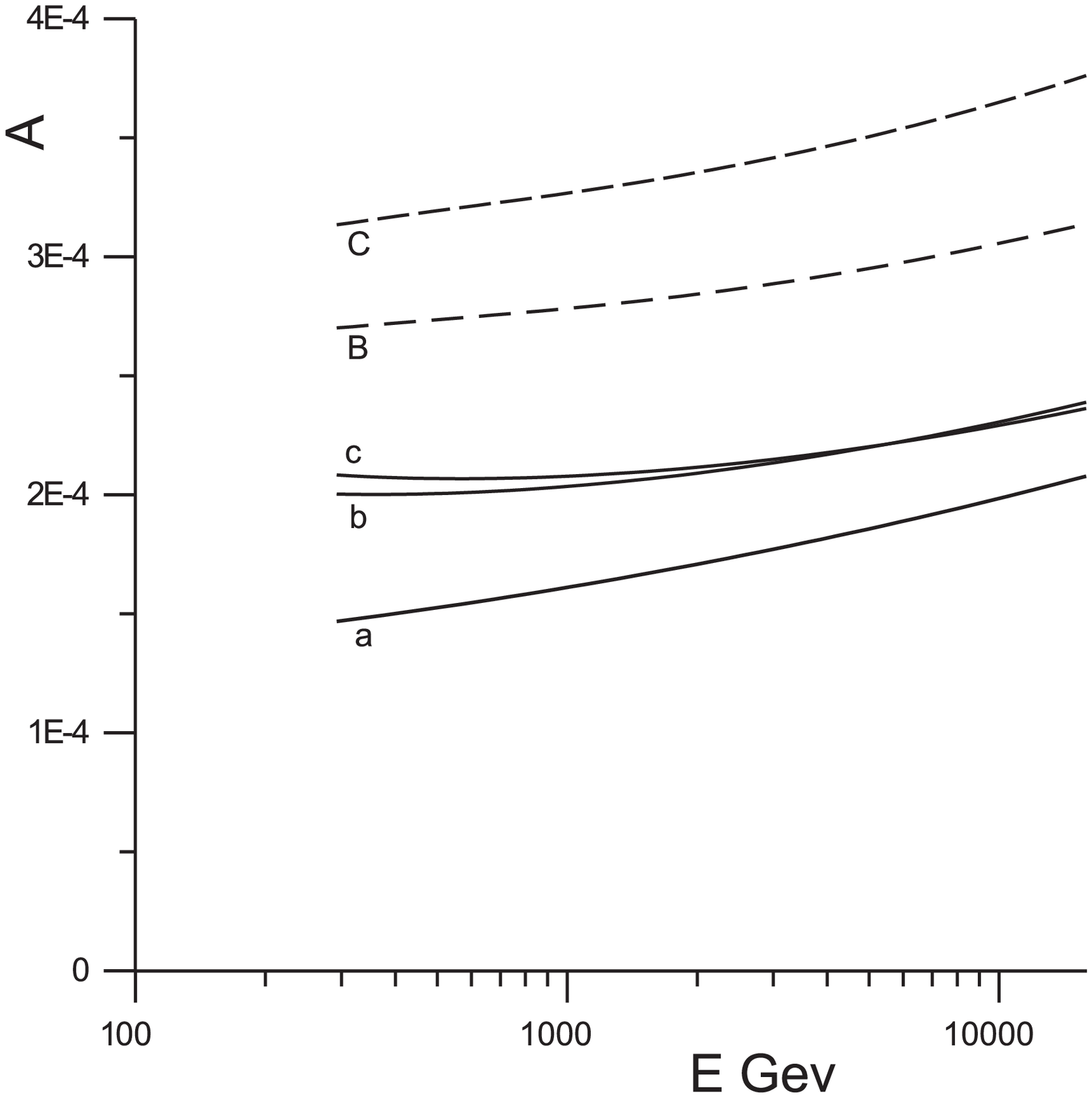}}
\caption{
Energy dependence of $\Omega^{\hbox{-}}$-hyperon
spin dichoism
in hydrogen target, calculated  for the
different form of  $\gamma_m (b)$ : curve (a) corresponds to the
 equation (23),
   (b,B) - (22), (c,C) -(20).
}
\end{figure}
\end{document}